\documentclass[journal]{IEEEtran}

\ifCLASSINFOpdf
  \usepackage[pdftex]{graphicx}
  \usepackage{caption2}

  \graphicspath{{../pdf/}{../jpeg/}}
   \DeclareGraphicsExtensions{.pdf,.jpeg,.png}
\else
  \usepackage[dvips]{graphicx}
  \graphicspath{{../eps/}}
  \DeclareGraphicsExtensions{.eps}
\fi
\usepackage{amsmath}
\begin{document}

%
\title{A Compact Pattern Reconfigurable Antenna Based on Multimode Plane Spiral OAM}


\author{\IEEEauthorblockN{Zelin Zhu,
Shilie Zheng,
Xiaowen Xiong,
Yuqi Chen,
Xiaofeng Jin, 
Xianbin Yu and
Xianmin Zhang}

\thanks{ 
Corresponding author: Shilie Zheng (email: zhengsl@zju.edu.cn).}}

%
%



\IEEEtitleabstractindextext{%
\begin{abstract}
   
Orbital angular momentum (OAM) has attracted great interest in the past few decades due to its helical phase characteristics. It has been proved that Plane spiral OAM (PSOAM) has potential in pattern reconfiguration of azimuthal domain because it solves the difficulty of mode superposition caused by different main lobe directions of different modes. However, the absence of a compact high-purity multimode PSOAM antenna hinders the development of PSOAM applications. In this paper, a compact high-purity 8-mode PSOAM antenna consisting of a coaxial resonator group (CRG) and a rotating parabolic reflector (RPR) is designed. It works in the X-band with the bandwidth of 140 MHz, and its efficiency is greater than 90$\%$. Amplitude and phase of each PSOAM beam generated by this antenna can be controlled independently owing to its feed network and separate resonators design, hence any target pattern in azimuthal domain can be approximately reconfigured by modes weighting. The simulation and experiment results show that this antenna can make the PSOAM based pattern configuration possible, which will have potential in the OAM based applications.
\end{abstract}

}

\maketitle

\IEEEdisplaynontitleabstractindextext

%
\IEEEpeerreviewmaketitle

\section{Introduction}
%
%
%
%

\IEEEPARstart{T}{hanks} to the helical phase characteristics and mode orthogonality, orbital angular momentum (OAM) has spawned many applications \cite{Allen1992Orbital}\cite{GattoFree}. For instance, OAM can improve spectrum utilization in wireless communication system \cite{Intrinsic}\cite{TamburiniEncoding}\cite{Zhang2016Orbital}, and it can also be used in rotational speed measurement and compression sensing imaging in radar system \cite{LaveryDetection}\cite{LiuOrbital}. With the development of the application research, multi-mode OAM wave is much more anticipated. Besides the OAM mode multiplexing, different OAM mode having different azimuthal phase distribution of $e^{-jl\varphi}$ offers a complete set of eigenmodes for electromagnetic waves in azimuthal domain, hence any target pattern in azimuthal domain can be easily reconfigured by the eigenmodes' weighting.
    
The traditional OAM-carrying beam has two limitations. One is that single mode OAM-carrying beam exists a central empty due to the phase uncertainty in the center, increasing the difficulty of long-distance signal reception \cite{ThidUtilization}. The other is that the beam carrying OAM of mode $l$ has an amplitude modulation, which is proportional to $J_{l}(kasin\theta)$, where $\theta$ is the zenith angle. This means that different mode $l$ has different main lobe direction (MLD) and this difference will seriously influence the mode superposing and the pattern reconfiguration.
 
Recently, the plane spiral OAM (PSOAM) waves have been proposed to solve these problems \cite{zzf2015Plane}. Comparing with the traditional OAM-carrying beam, PSOAM beam is end-fire and omnidirectional in azimuthal domain. In this case, all PSOAM beams with different modes have the same MLD in zenith direction, so the mode diversity in azimuthal domain can be well utilized to achieve pattern reconfiguration \cite{Zheng2018Realization}. 

Pattern reconfigurable antenna is always a research hotspot, as this kind of antenna can provide a flexible beam according to the operation requirements \cite{Godara2004Smart}. It can also be used as a spatial coding scheme \cite{cui2019direct}. Usually, phased array antenna is used. Compared phase array antenna with the PSOAM based pattern reconfiguration, the latter can easily configure any pattern just by calculating the fast Fourier transform (FFT) of target pattern, and then feeding each mode with the corresponding amplitude and phase. 

\begin{figure}[t]
 \centering
 \includegraphics[width=3.0in]{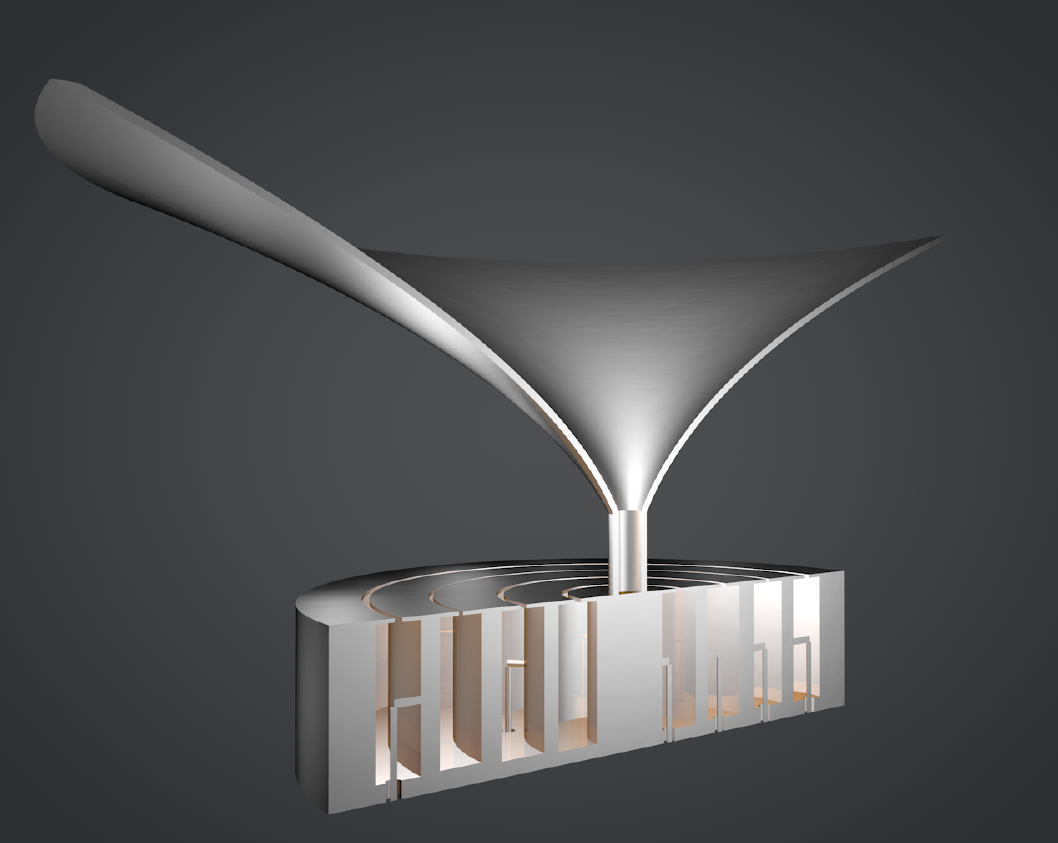}
 \caption{Cross section view of the antenna model}
 \label{fig1}
\end{figure}

Up to now, the antenna which can generate PSOAM beam includes circular ring resonator antenna \cite{zzf2015Plane}, array antenna \cite{maqing}, dielectric coaxial resonator antenna \cite{dongruofan}, antenna with meta-surface \cite{hualin}, and so on. No matter which method is used, the generated PSOAM modes are no more than 4, which limits the applications of multimode OAM-carrying beam. In 2018, G. Junkin used TE coaxial resonators and a TM coaxial resonator to generate two modes OAM beams by adjusting the size of TM resonator to make the extreme points of different order Bessel functions concurrently simulated \cite{junkinsolo}. This scheme demonstrated the advantages of TE coaxial resonator in designing multimode OAM antenna. However it lacks the measured results and the impedance match was difficult.

  In this paper, a compact multimode PSOAM antenna operating in the X-band with the bandwidth of 140 MHz is proposed. It consists of a coaxial resonator group (CRG) and a rotating parabolic reflector (RPR). Thanks to the separation of CRG and RPR, this antenna can also be used as a compact eight-mode OAM antenna without RPR. Since the amplitude and phase of each generated PSOAM beam can be controlled independently, various target patterns in azimuthal domain can be approximately reconfigured by the eight modes' weighting. Especially, the pencil beam can be easily constructed and its azimuthal beam scanning can be realized by feeding each mode with the same amplitude and regular phase.
  
This paper is organized as follows: Section \uppercase\expandafter{\romannumeral2} presents theoretical analysis of the field distribution in the resonator and the farfield radiation of the proposed antenna. For higher accuracy, the coaxial resonator model is used instead of the head-tail-jointed rectangular waveguide model to analyze the resonator field distribution. Vector diffraction theory is used to calculate the reflection field of RPR and the radiation field of this antenna. Section \uppercase\expandafter{\romannumeral3} gives the detailed design of this antenna, including the antenna components and the feed network. Section \uppercase\expandafter{\romannumeral4} shows the simulated and measured results of both the antenna performance and its capability in the pattern reconfiguration. Finally, the pros and cons of this antenna is discussed from the practical point of view for the future application.

\vspace*{-1.5mm}
\section{theoretical analysis}

\subsection{The Coaxial Resonator Model}

The main part of the proposed OAM antenna is an integrated coaxial ring resonator. Its field distribution is anlalyzed by a strict coaxial resonator model. Comparing with that in \cite{ZhengTransmission}, this field analysis will be more exact. It considers the bending of sidewalls of the waveguide. The different boundary conditions lead to the difference of field distribution inside the resonator, and thus affect the radiation field from the slot.

 \begin{figure}[htb]
 \centering
 \includegraphics[width=3.0in]{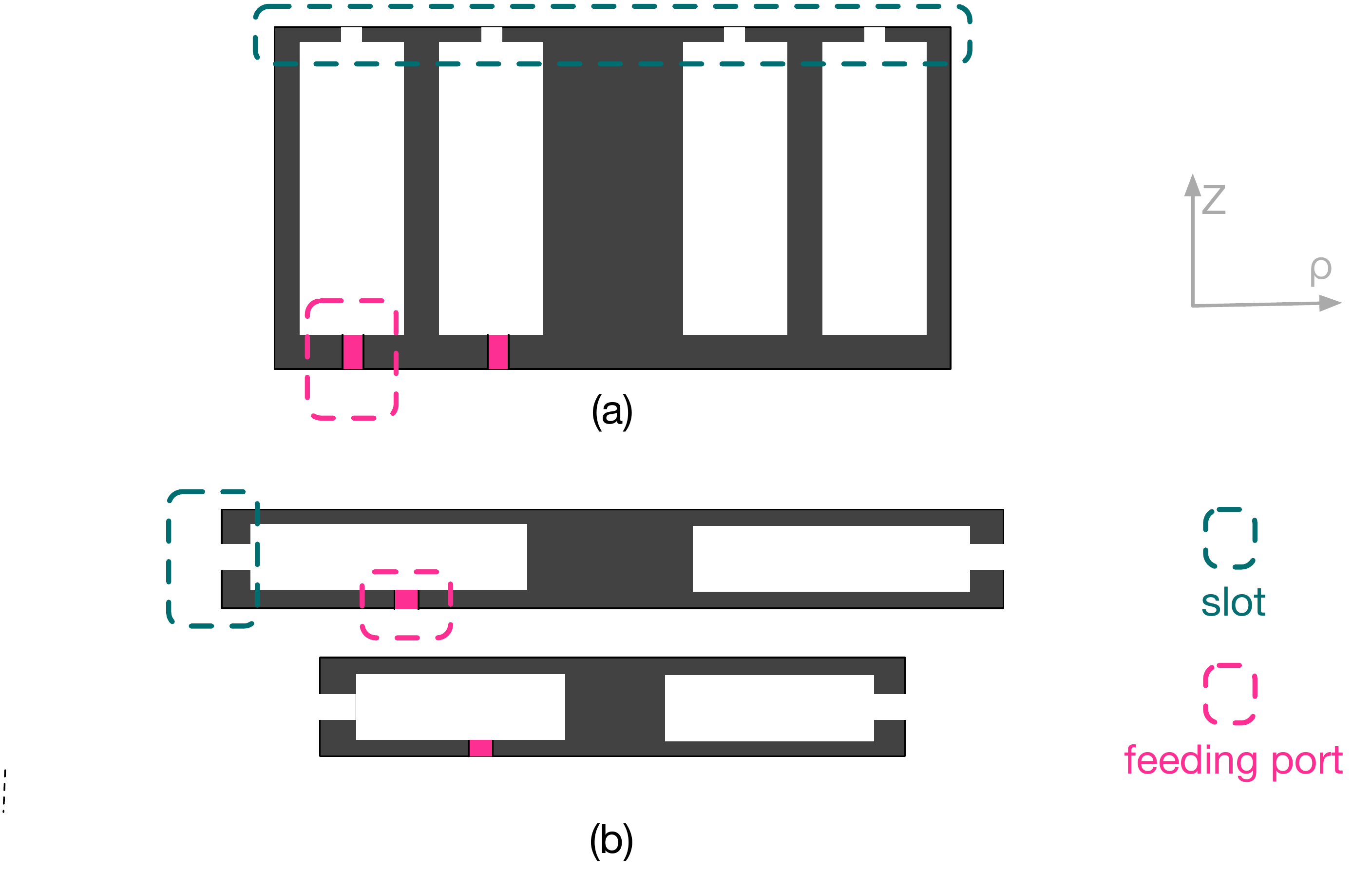}
 \caption{Cross section view of two kinds of coaxial resonator models: (a) top slotted TE coaxial resonator; (b) sidewall slotted TM coaxial resonator }
 \label{duibi}
\end{figure}

The coaxial resonator radiation can be divided into two cases: A) top slotted TE coaxial resonator; B) sidewall slotted TM coaxial resonator, as shown in Fig.\ref{duibi}. When multimode radiation is anticipated, multiple resonators with different radius can be radially arranged for case A or vertically stacked for case B. It is obvious that the feeding ports will be obstructed by the shell of the adjacent resonators in case B. Hence case A is selected to design the CRG to realize multi-mode OAM generation.

To generate the OAM wave of mode $l$, $TE_{l11}$ mode should be stimulated. According to its Borgnis function shown in Eq.\ref{q}, the magnetic field at each point of slot will have a radial component $\vec H_{\rho}$ and an azimuthal component $\vec H_{\varphi}$ \cite{zhangkeqian}. 
\begin{equation}
\begin{split}
\label{q}
 U(\rho, \varphi, z)&=U_{0}\lbrack N^{'}_{l}(TR_{ex})J_{l}(T\rho)-J^{'}_{l}(TR_{ex})N_{l}(T\rho)\rbrack \\ &cos(l\varphi)cos(\beta z)
 \end{split}
\end{equation}

Using the two-ports feed method of traditional model based on the Euler formula, azimuthal standing wave changes into traveling wave \cite{ZhengTransmission}. In this case, the two magnetic components can be described as 
\begin{equation}
\begin{split}
\label{hp}
  &H_{\varphi}(\rho,\varphi) =\frac{2\beta lU_{0}}{\rho}\\
  &\lbrack N^{'}_{l}(TR_{ex})J^{'}_{l}(T\rho)-J^{'}_{l}(TR_{ex})N^{'}_{l}((T\rho)\rbrack e^{-jl\varphi}
  \end{split} 
\end{equation}
\begin{equation}
\begin{split}
\label{hr}
  &H_{\rho}(\rho,\varphi) =-2j\beta TU_{0}\\
 &\lbrack N^{'}_{l}(TR_{ex})J^{'}_{l}(T\rho)-J^{'}_{l}(TR_{ex})N^{'}_{l}((T\rho)\rbrack e^{-jl\varphi}
  \end{split} 
\end{equation}
where $R_{ex}$ is the external radius, $\omega_{c}=2\pi f_{c}$ is the angular frequency and $T=\omega_{c}\sqrt{\mu\varepsilon}$ is the eigenvalue. The corresponding equivalent current source at each point of slot $\vec Js$ can be expressed as  
 \begin{equation}
 \begin{split}
 \label{js}
 \vec Js(\rho,\varphi) &=-\vec e_{z}\times \vec H=H_{\varphi}(\rho,\varphi)\vec e_{\rho}+H_{\rho}(\rho,\varphi)\vec e_{\varphi}
 \end{split}
\end{equation}

\begin{figure}[htb]
 \centering
 \includegraphics[width=3.0in]{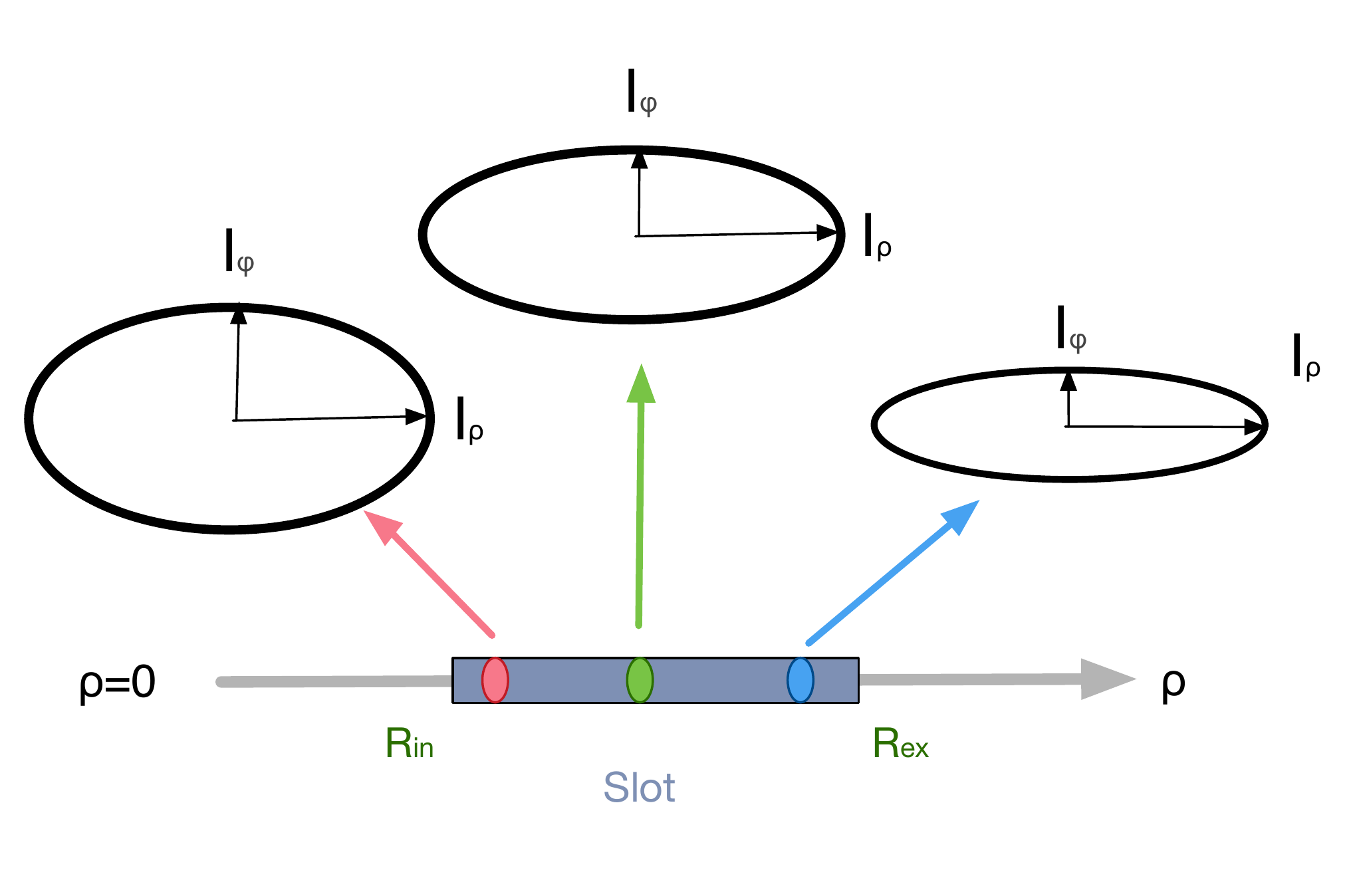}
 \caption{Current composition at different radial positions}
 \label{dianliu}
\end{figure}

 In  \cite{ZhengTransmission}, the slot is thought to have a homogeneous radial ring surface current $\vec I=\vec I_{\rho}e^{-jl\varphi}$, where $\vec I(\cdot)$ is directly proportional to $\vec Js(\cdot)$. Actually, there have two orthogonal currents $\vec I_{\varphi}(\rho)$ and $\vec I_{\rho}(\rho)$ with a $90^{o}$ phase difference at each point between the internal radius $R_{in}$ and external radius $R_{ex}$ of the slot. As shown in Fig.\ref{dianliu}, the currents at three points marked red, blue and green color between the two edges, have different ratio of the two orthogonal currents component, which are related to its radial distance and can be calculated directly according to Eq.\ref{hp} and Eq.\ref{hr}.  

\begin{figure}[htb]
 \centering
 \includegraphics[width=3.0in]{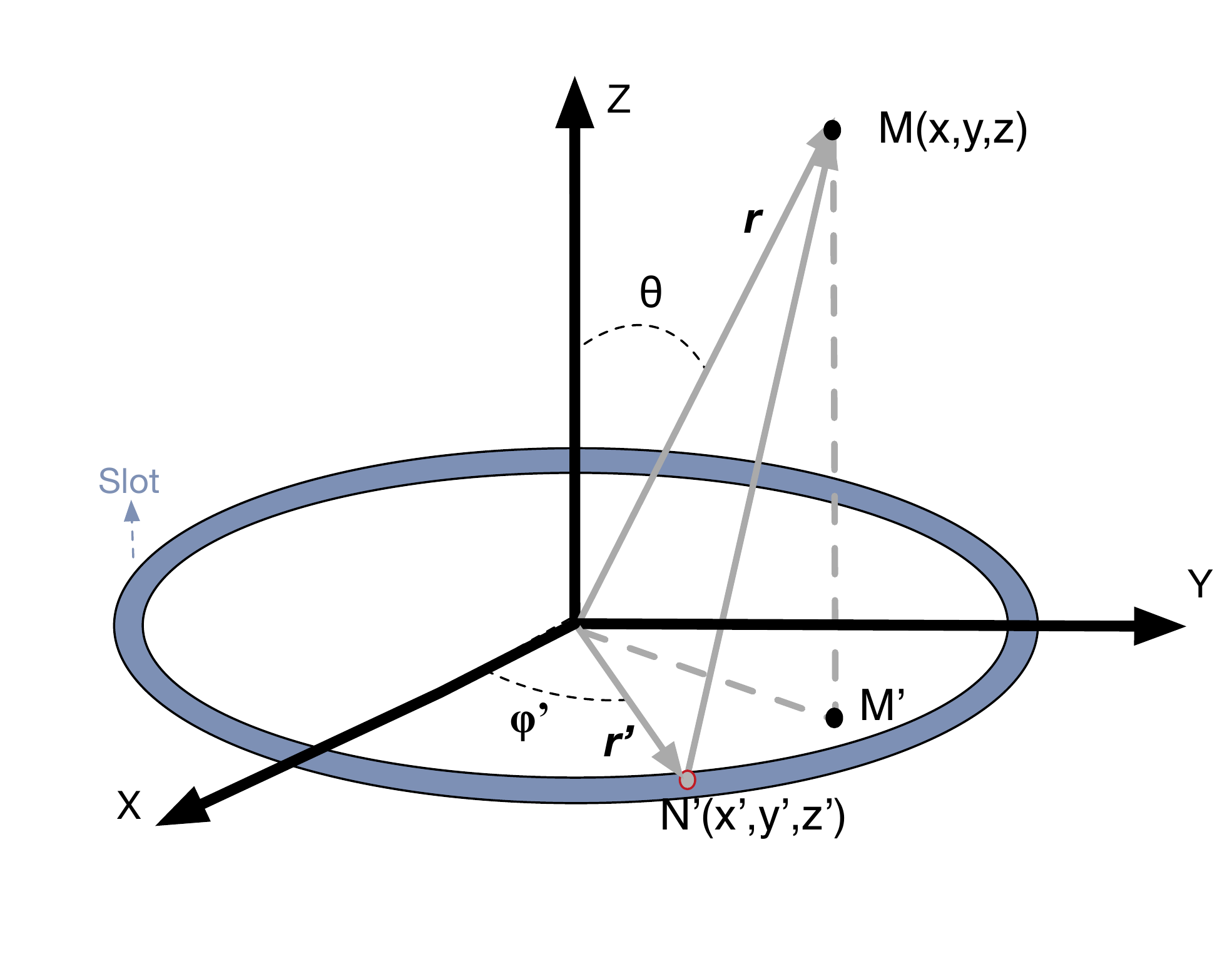}
 \caption{Modified radiation field integral model}
 \label{zuobiao}
\end{figure}

Fig.\ref{zuobiao} shows the modified radiation field integral model of the circular slot area, $N^{'}(x^{'},y^{'},z^{'})$ is any point in it. $M(x,y,z)$ is the calculated point. In this case, the vector potential equation of the complex current source can be described as  
 \begin{equation}
 \label{is}
 \vec A(\vec r)=\frac{\mu_{0}}{4\pi}\int^{R_{ex}}_{R_{in}}\int^{2\pi}_{0}\frac{\vec I(r^{'})e^{-jl\varphi}e^{-jk|\vec r-\vec r^{'}|}}{|\vec r-\vec r^{'}|}  r^{'}d\varphi^{'} dr^{'}
\end{equation}
  
 Here, $I(r^{'})$ is the combination of the two orthogonal currents based on the above analysis. The radiation field is a vector superposition of the two current components.

  \subsection{Vector Diffraction Theory}
In order to adjust each OAM beam's MLD to $90^{o}$, RPR is adopted in this proposed antenna. Since the antenna aperture and the RPR is of the same order of magnitude as the wavelength, the Stratton-Chu (S-C) formula based on the strict vector diffraction theory is more accurate to calculate the radiation and reflection fields than geometrical optics theory \cite{scfunction}. For the radiation field of the antenna, S-C formulas can be expressed as
  \begin{equation}
   \begin{split}
 \label{sce}
 \vec E^{'}&=\int_{S}\lbrace j\omega\mu\lbrack \vec n\times \vec H(\vec r)\rbrack g(\vec r,\vec r^{'}) \\
 &+\lbrack \vec n \cdot \vec E(\vec r)\rbrack\nabla g(\vec r,\vec r^{'})+\lbrack \vec n \times \vec E(\vec r)\rbrack \times \nabla g(\vec r,\vec r^{'})\rbrace dS 
\end{split}
\end{equation}
\begin{equation}
   \begin{split}
 \label{scf}
 \vec H^{'}&=\int_{S}\lbrace -j\omega\varepsilon\lbrack \vec n\times \vec E(\vec r)\rbrack g(\vec r,\vec r^{'}) \\
 &+\lbrack \vec n \cdot \vec H(\vec r)\rbrack\nabla g(\vec r,\vec r^{'})+\lbrack \vec n \times \vec E(\vec r)\rbrack \times \nabla g(\vec r,\vec r^{'})\rbrace dS 
\end{split}
\end{equation}
where $S$ is the radiation source area, $\vec n$ is the outer normal of $S$ surface, and $g$ is the space Green function. For the reflection field from RPR, the S-C formula can be simplified as
 \begin{equation}
   \begin{split}
 \label{sc2e}
 \vec E^{'}&=-\frac{2j}{\omega\varepsilon}\int_{S}\lbrace k^{2}\lbrack \vec n\times \vec H(\vec r)\rbrack g(\vec r,\vec r^{'}) \\
 &+\lbrack \vec n \times \vec H(\vec r)\rbrack \cdot \nabla\lbrack \nabla g(\vec r,\vec r^{'})\rbrack\rbrace dS 
\end{split}
\end{equation}
\begin{equation}
   \begin{split}
 \label{sc2f}
 \vec H^{'}&=\int_{S}2\lbrack \vec n \times \vec H(\vec r)\rbrack \times \nabla g(\vec r,\vec r^{'}) dS 
\end{split}
\end{equation}

 The above four formulas can be used to get the expression of each field component in different coordinate systems through coordinate transformation, which is convenient for numerical simulation of the field distribution at any position in the near or far fields.

\begin{figure}[htb]
 \centering
 \includegraphics[width=3.0in]{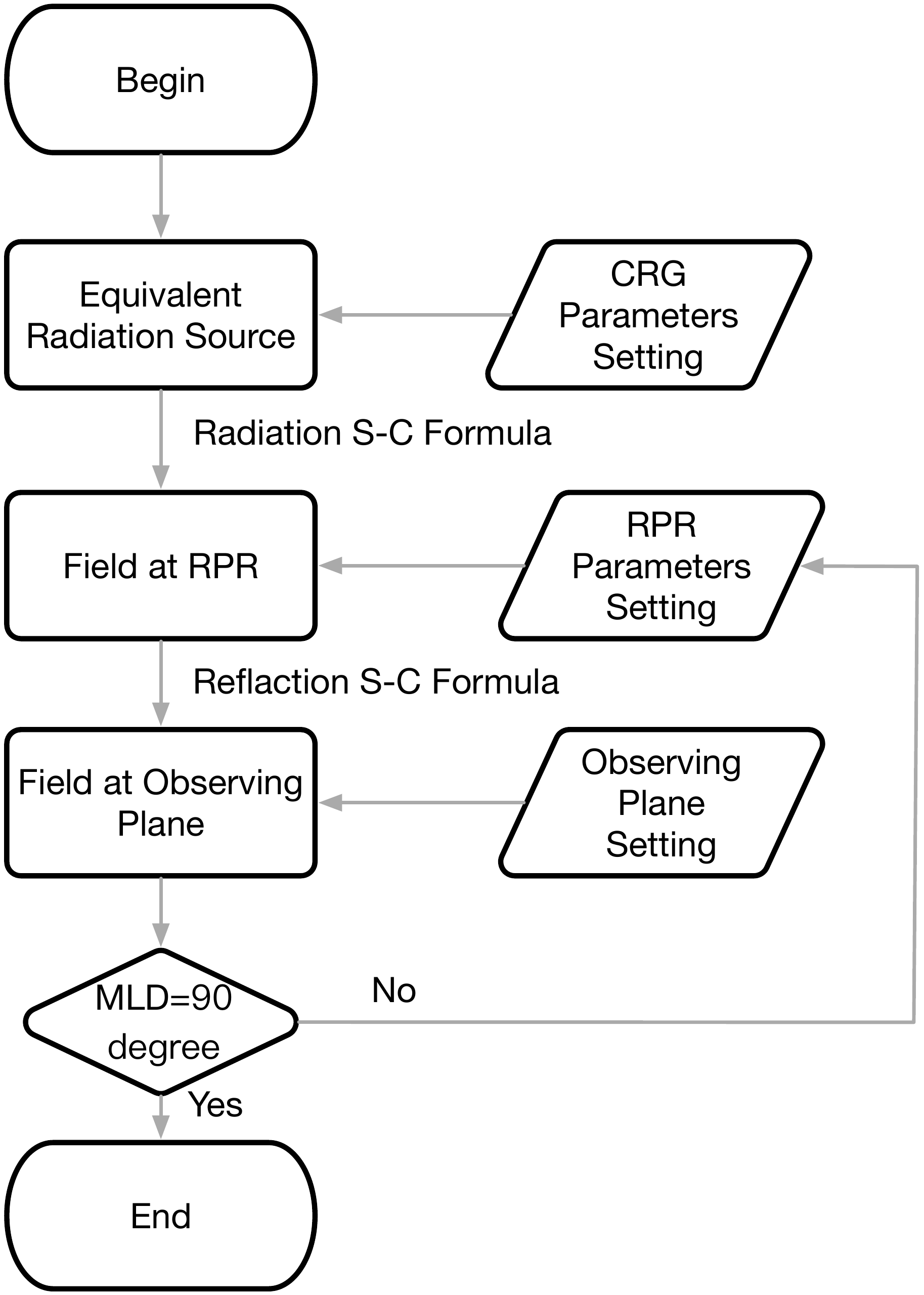}
 \caption{ Flow Chart of RPR Design Using S-C formulas}
 \label{liucheng}
\end{figure}

  Fig.\ref{liucheng} shows the flow chart of RPR design using four S-C formulas. First of all, the parameters of the CRG, the RPR and the observing plane at $\varphi=0^{o}$ should be set. Secondly, the equivalent radiation source at the slots of CRG can be expressed according to Eq.\ref{hp} and Eq.\ref{hr}. Thirdly, the field at the the surface of the RPR radiated from the slots of the CRG can be calculated using the radiation S-C formula Eq.\ref{sce} and Eq.\ref{scf}, and then the field at the observing plane reflected for the surface of RPR can be calculated using the reflection S-C formula Eq.\ref{sc2e} and Eq.\ref{sc2f}. Lastly, keeping adjusting the focal length of the parabola of the RPR according to the MLD of the field at the observing plane, until the MLD points to $\theta=90^{o}$.
  
\section{The design of antenna}

\subsection{Antenna}
\begin{figure}[htb]
 \centering
 \includegraphics[width=3.0in]{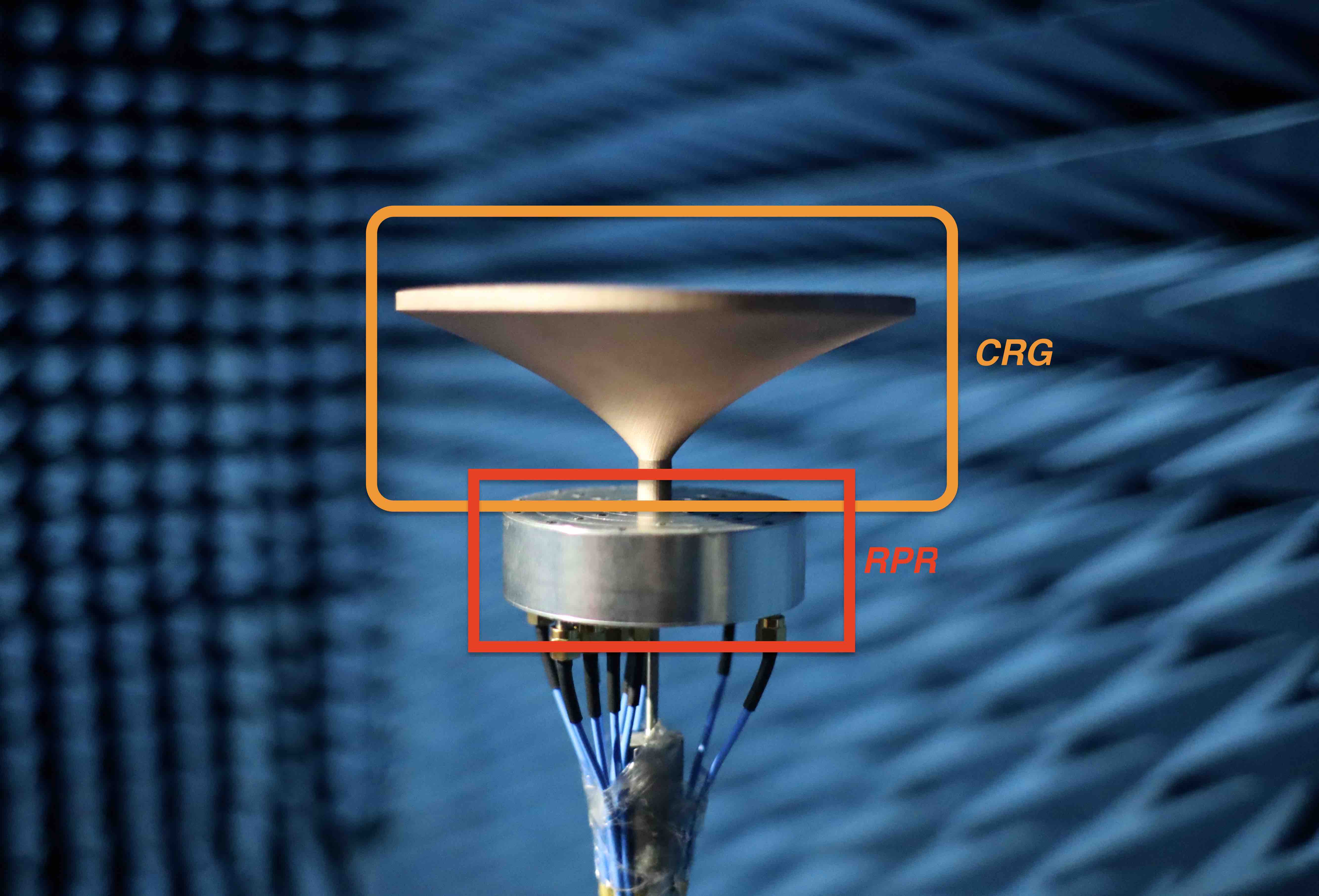}
 \caption{ Front view of Antenna}
 \label{shiwu}
\end{figure}

\begin{figure}[htb]
 \centering
 \includegraphics[width=3.0in]{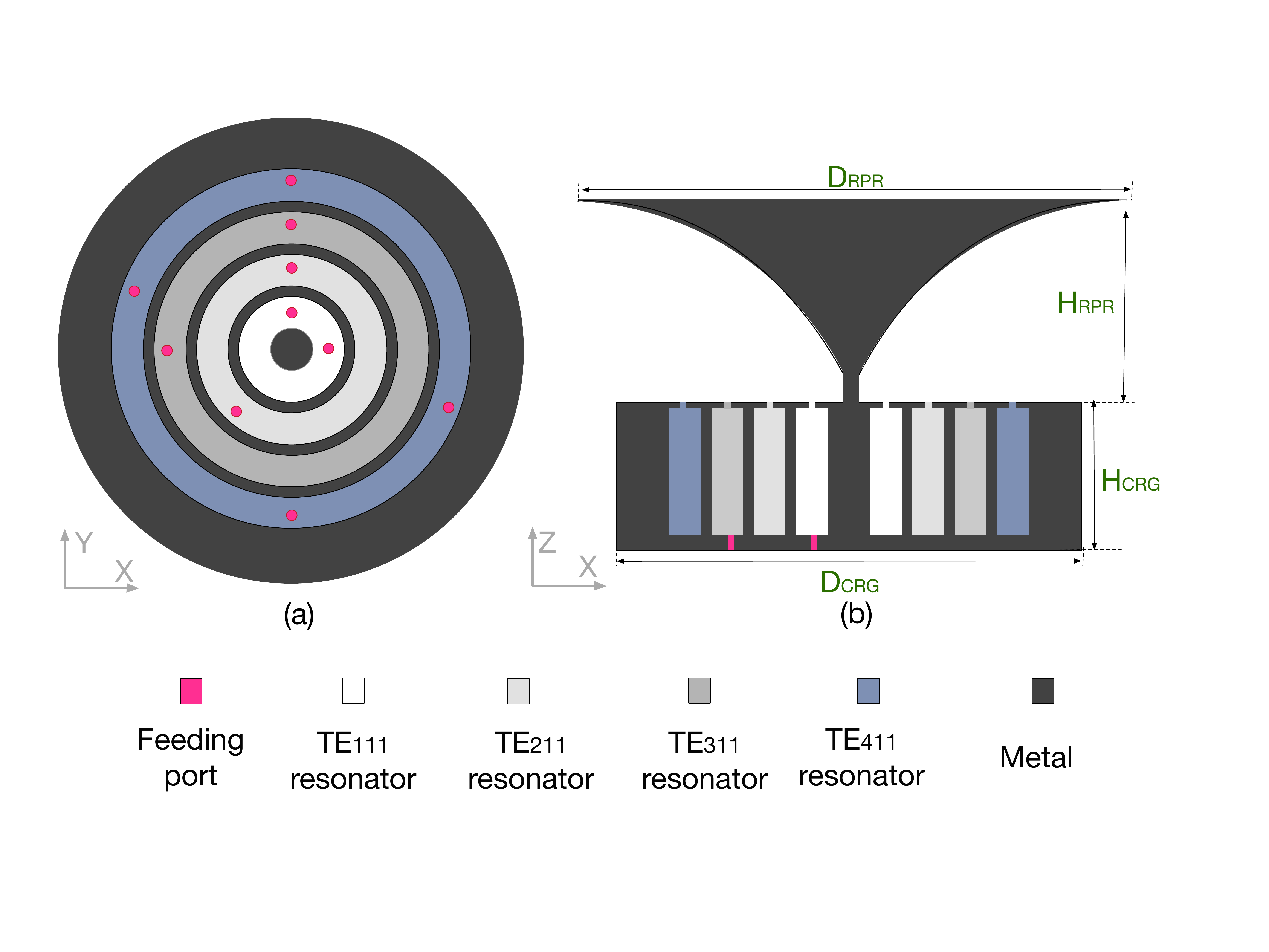}
 \caption{Cross section view of Antenna model(a) XOY plane cross section view of CRG (b) XOZ plane cross section view of the whole antenna}
 \label{jiegou}
\end{figure}

 The antenna consisting of a CRG and a RPR is shown in Fig.\ref{shiwu}, and its cross section view is shown in Fig.\ref{jiegou}. The CRG  whose diameter is $D_{CRG}=90.0$ $mm$ and height is $H_{CRG}=24.0$ $mm$ consists of four top slotted TE coaxial resonators sequentially arranged from inside to outside. The dimensions of the four resonators are shown in the Tab.\ref{t-1}. They work in $TE_{l11}$ mode, $l=1,2,3,4$ respectively.
 
 \begin{table}[htb]
 \setlength{\abovecaptionskip}{0.cm}
\setlength{\belowcaptionskip}{0.3cm}
\centering
\caption{THE DIMENSIONS OF THE FOUR RESONATORS}
\label{t-1}       
\begin{tabular}{l|llll}
\hline\noalign{\smallskip}
$Mode$ &$TE_{111}$ &$TE_{211}$  &$TE_{311}$  &$TE_{411}$   \\
\noalign{\smallskip}\hline\noalign{\smallskip}
Inside radius& 5.9mm&14.6mm& 24.3mm & 33.0mm \\
External radius& 12.3mm & 21.2mm& 30.3mm & 39.0mm\\
Height& 19.20mm& 19.00mm& 18.60mm & 18.80mm \\
\noalign{\smallskip}\hline
\end{tabular}
\end{table}
Using two ports of the 90 degree hybrid, each resonator can generate two OAM waves with opposite sign independently. Hence the OAM beams of mode $l = \pm 1,\pm 2,\pm 3,\pm 4$ can be generated. For mode $\pm 4$, two pairs of the feeding ports are adopted to improve the amplitude uniformity.

 The RPR's diameter is $D_{RPR}=150$ $mm$ and height is $H_{RPR}=56$ $mm$. Its surface satisfies parabolic equation $Z=\sqrt{2*P*\rho}$, where the parabola focal length $P=21$ $mm$.
 
\subsection{Amplitude Phase Control Feed Network}
\begin{figure}[htb]
 \centering
 \includegraphics[width=3.0in]{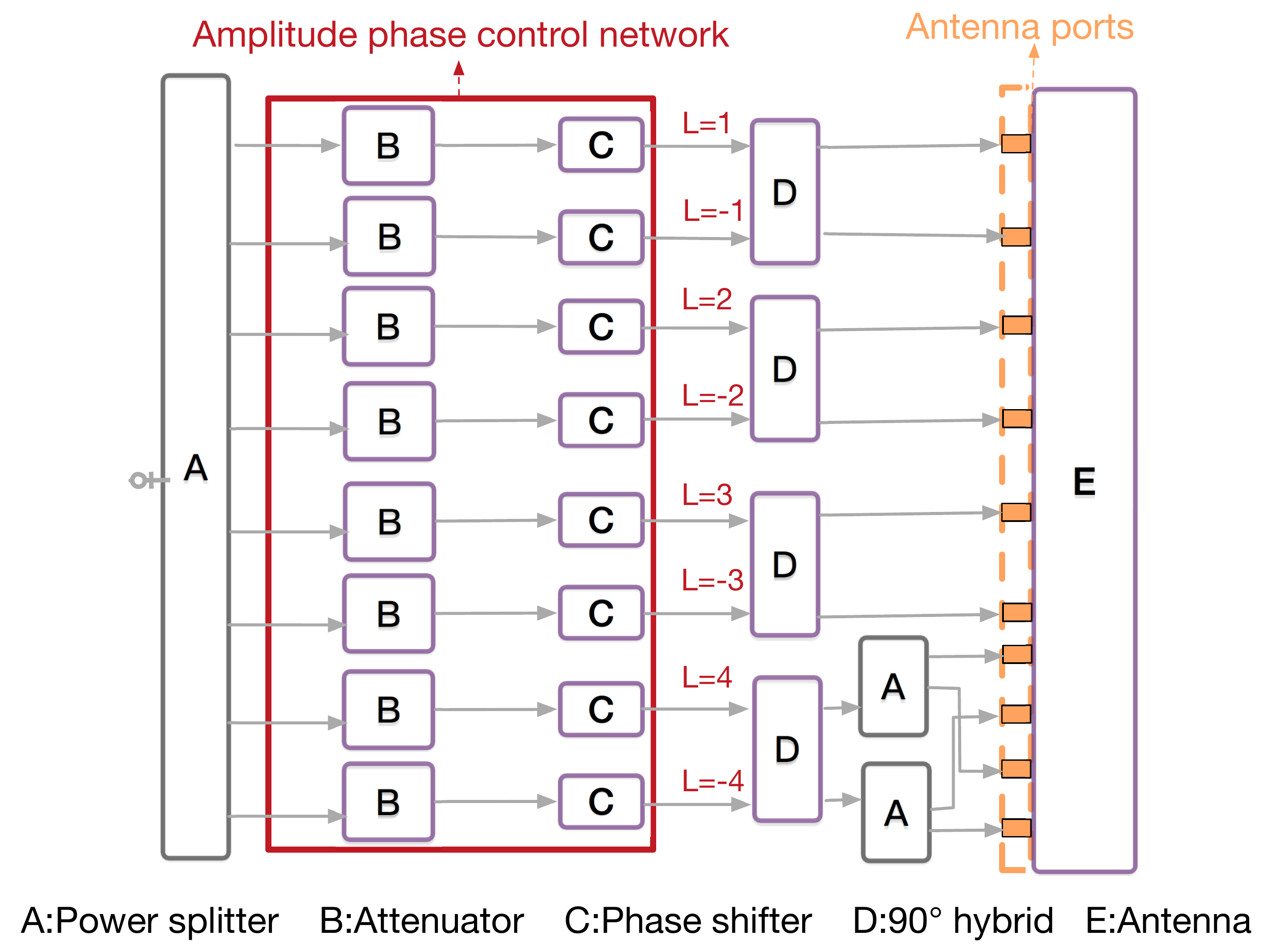}
 \caption{Feed network model}
 \label{net}
\end{figure}

\begin{figure}[htb]
 \centering
 \includegraphics[width=3.0in]{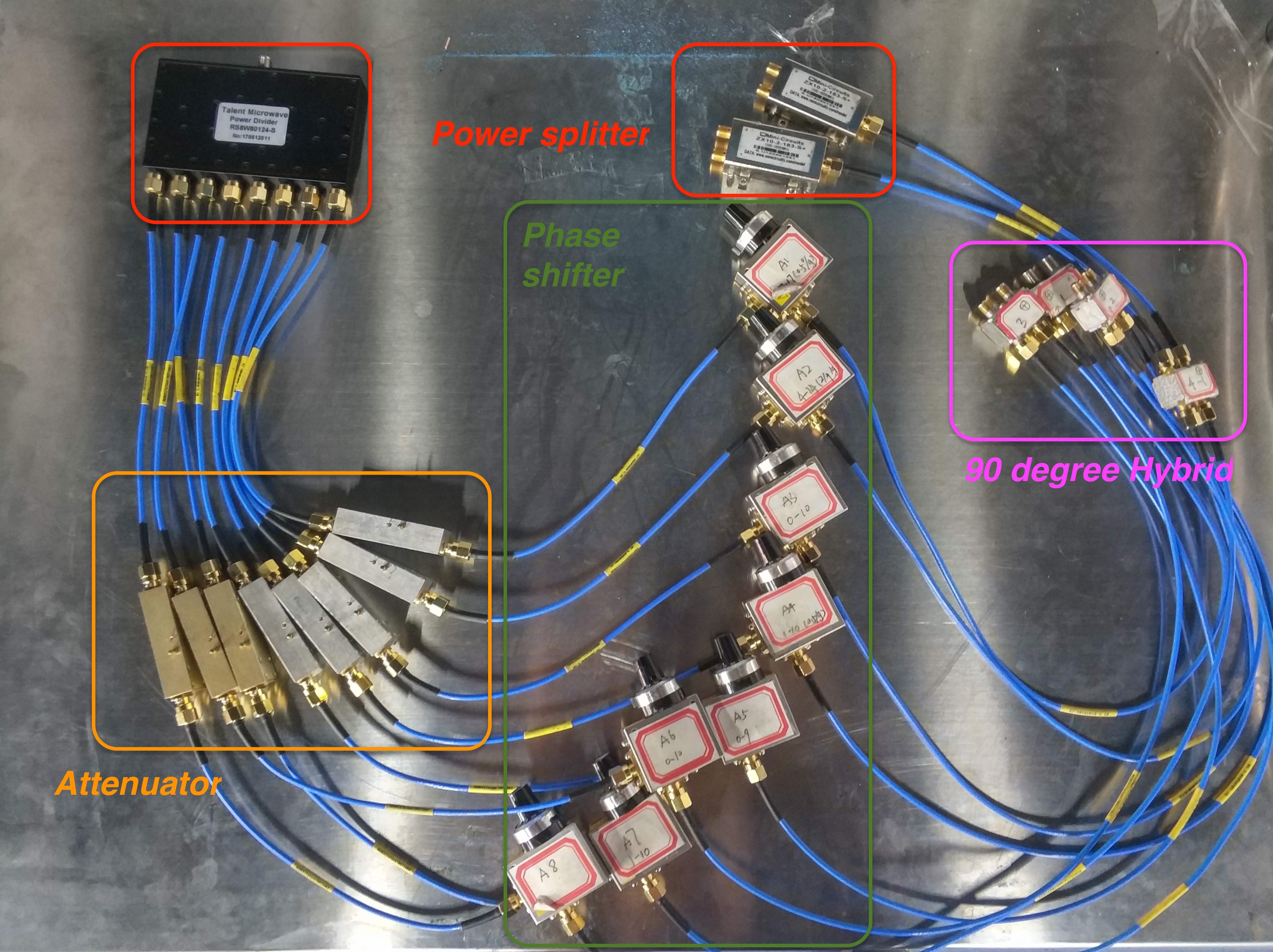}
 \caption{Photo of the feed}
 \label{net2}
\end{figure}

To realize the flexible pattern recomfiguration, the feed network system is carefully designed. Fig.\ref{net} and Fig.\ref{net2} show the model and the real configuration of the feed network, respectively.  

A one-to-eight power splitter is used to divide the RF signal into eight parts, each of them is followed by an attenuator and a phase shifter so that the eight port of the four 90 degree hybrids can be controlled separately, and thus each generated PSOAM can be controlled accordingly.
\section{The Simulation and Measurement of Antenna}
\subsection{Antenna Performance}
\begin{figure}[htb]
 \centering
 \includegraphics[width=3.0in]{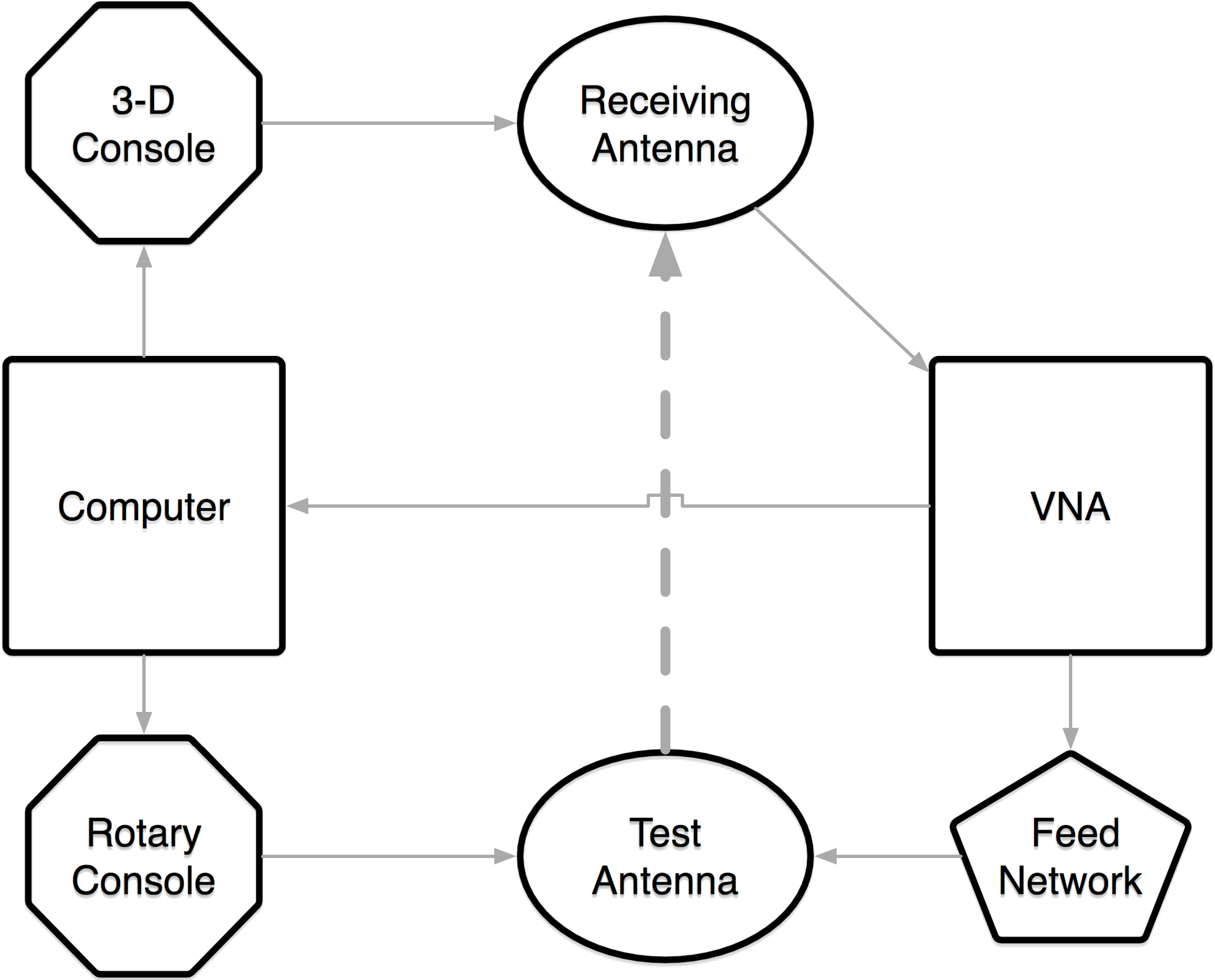}
 \caption{The setup of the measurement system}
 \label{net3}
\end{figure}

The setup of the measurement system is shown in Fig.\ref{net3}. The system includes: vector network analyzer(VNA), feed network system, three-dimensional console, rotary console and receiving antenna.

\begin{figure}[htb]
 \centering
 \includegraphics[width=3.0in]{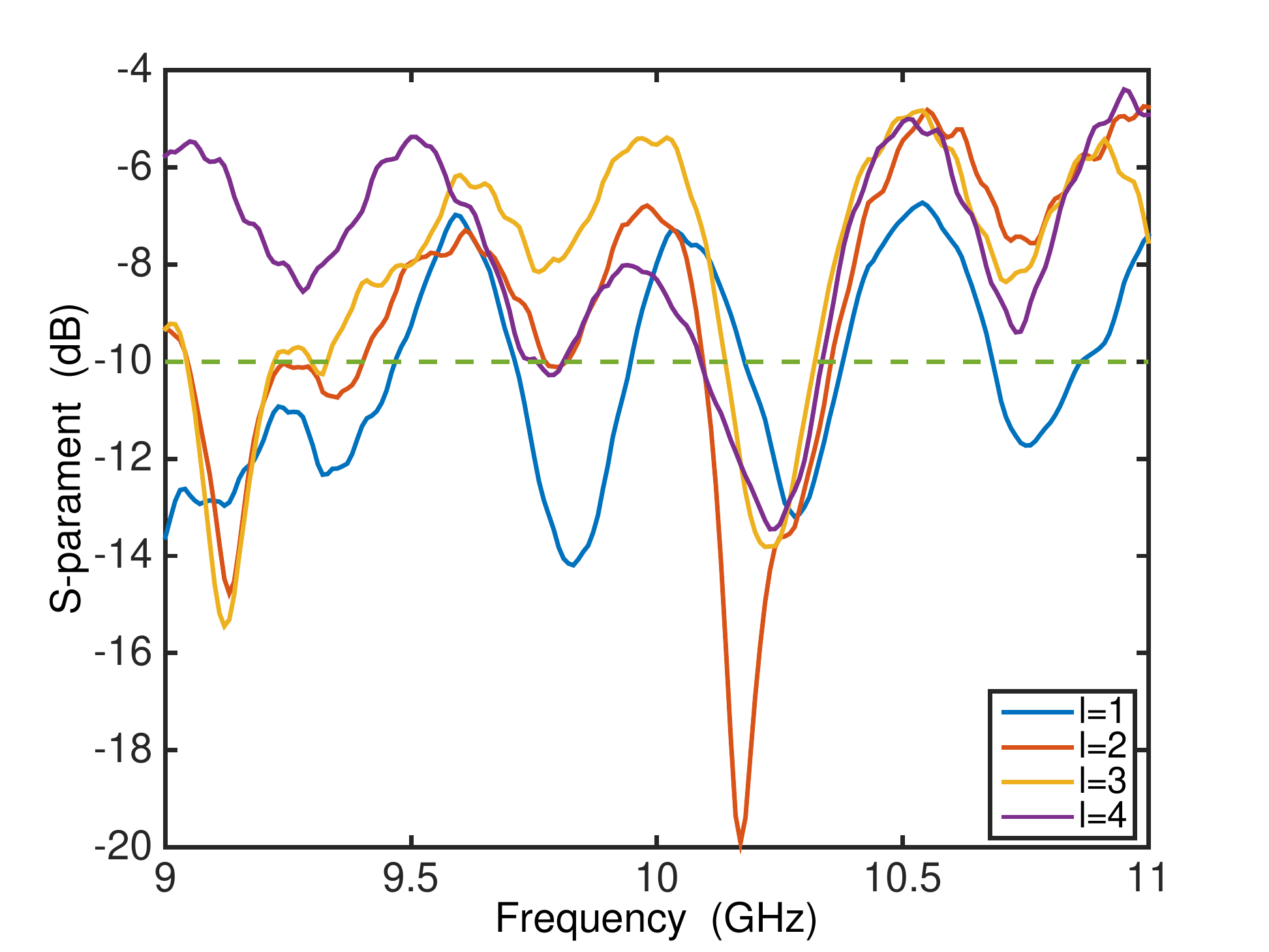}
 \caption{S-paraments measured results of single mode}
 \label{sp}
\end{figure}

Fig.\ref{sp} shows the measured S parameter for four PSOAM modes of $+1$,$+2$,$+3$ and $+4$. As mode $+l$ and $-l$ are symmetric, the results of the negative OAM mode are not shown. It can be seen that the working bandwidth of all the resonator are nearly the same, so the pattern reconfiguration can be realized in this frequency band. The center frequency $f_{s}=10.27$ $GHz$, and the total bandwidth is nearly $BW=140$ $MHz$.

\begin{figure}[htb]
 \centering
 \includegraphics[width=3.5in]{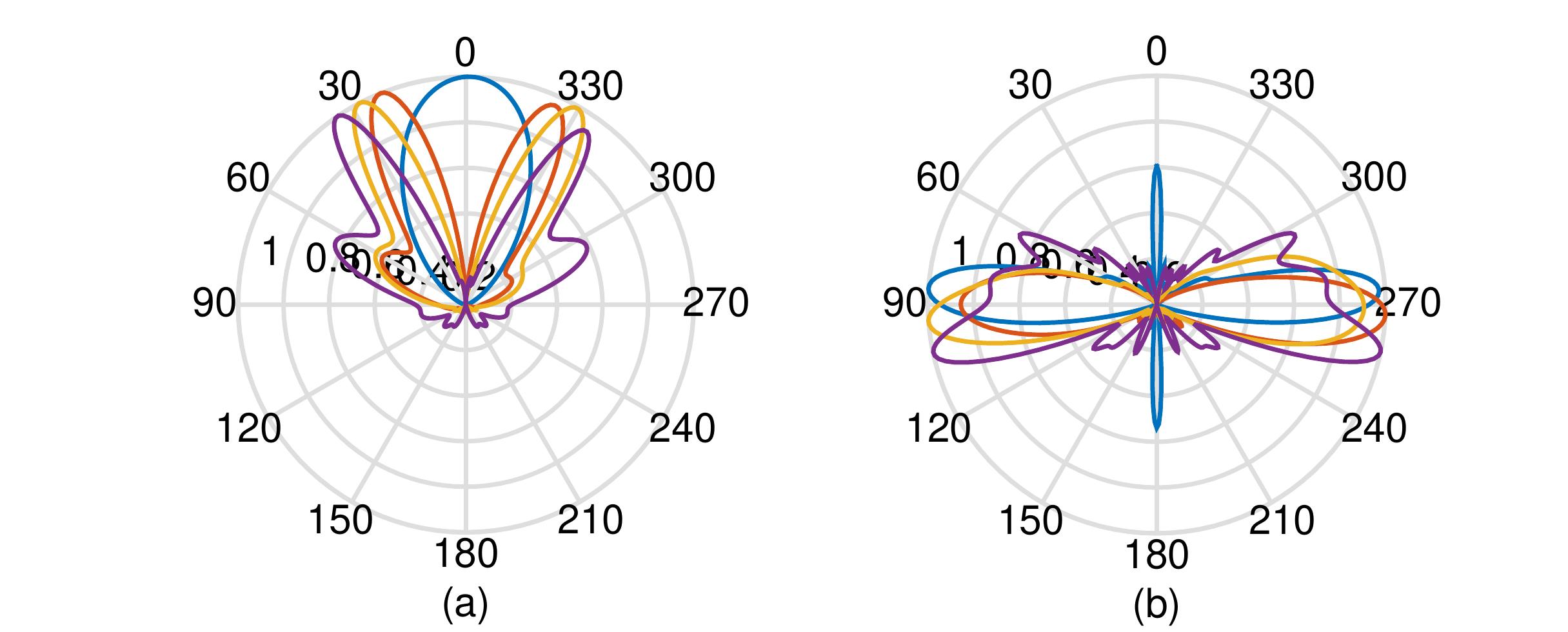}
 \caption{Simulated pattern at $\varphi=0^{o}$ plane: (a) without RPR (b) with RPR}
 \label{simu4}
\end{figure}

To demonstrate the changes brought by RPR, Fig.\ref{simu4} shows the simulated results of the MLD of each mode with and without RPR at $\varphi=0^{o}$ plane. The MLDs of the generated PSOAM waves are all around $\theta=90^{o}$. The main lobe width of the PSOAM waves are about $35^{o}-60^{o}$. Obviously, the PRR helps the OAM waves transform into PSOAM waves.

Tab.\ref{t0} shows the simulated and measured results of MLD with RPR and the relative power (RP) value in the direction of $\theta=90^{o}$. The definition of RP is the quotient of energy at $\theta=90^{o}$ plane and energy in MLD plane. It is displayed in dB coordinates, representing the power deviation in $\theta=90^{o}$ plane and MLD plane. The simulated and measured results are matched well. The MLDs of the generated PSOAM beams turn to $\theta=90^{o}$ nearby. The lower RP value means that even if the main lobe is not exactly at $\theta=90^{o}$, the energy at $\theta=90^{o}$ plane is close to the main lobe energy. Hence, the power efficiency is high and the superposition at $\theta=90^{o}$ is feasible.

\begin{figure}[htb]
 \centering
 \includegraphics[width=3.0in]{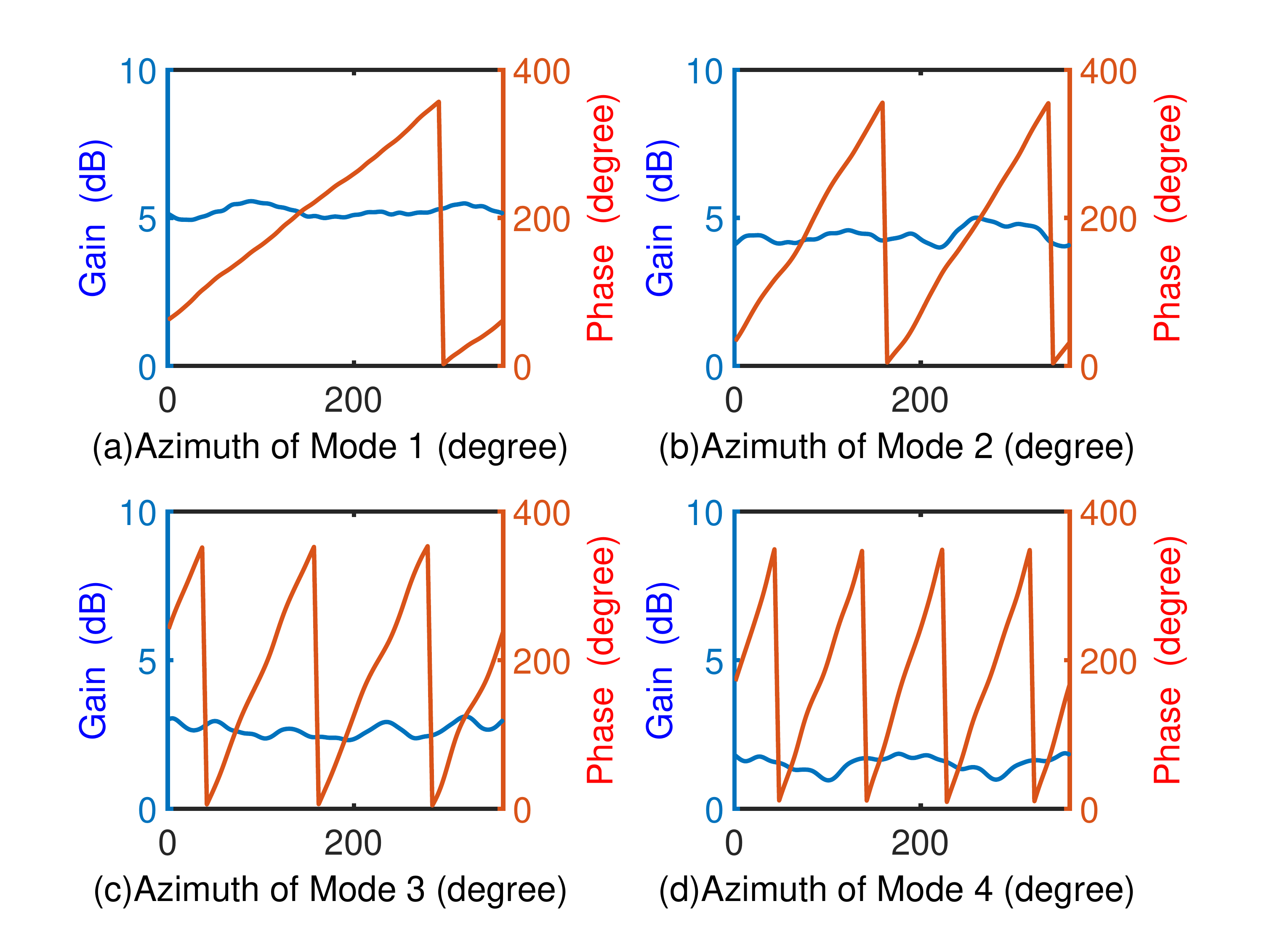}
 \caption{Gain and phase distribution of single mode at the position of $\theta=90^{o}$ (Simu): (a) $l=1$ (b) $l=2$ (c) $l=3$ (d) $l=4$}
 \label{single}
\end{figure}

In order to analyze the vortex characteristic of the field, Fig.\ref{single} shows the simulated results of the amplitude and phase distribution of each mode at $\theta=90^{o}$. The simulated results show that single mode PSOAM beam has good phase linearity and amplitude uniformity in the azimuthal domain. With the increasing of the OAM mode, the uniformity of amplitude decreases.

\begin{figure}[htb]
 \centering
 \includegraphics[width=3.0in]{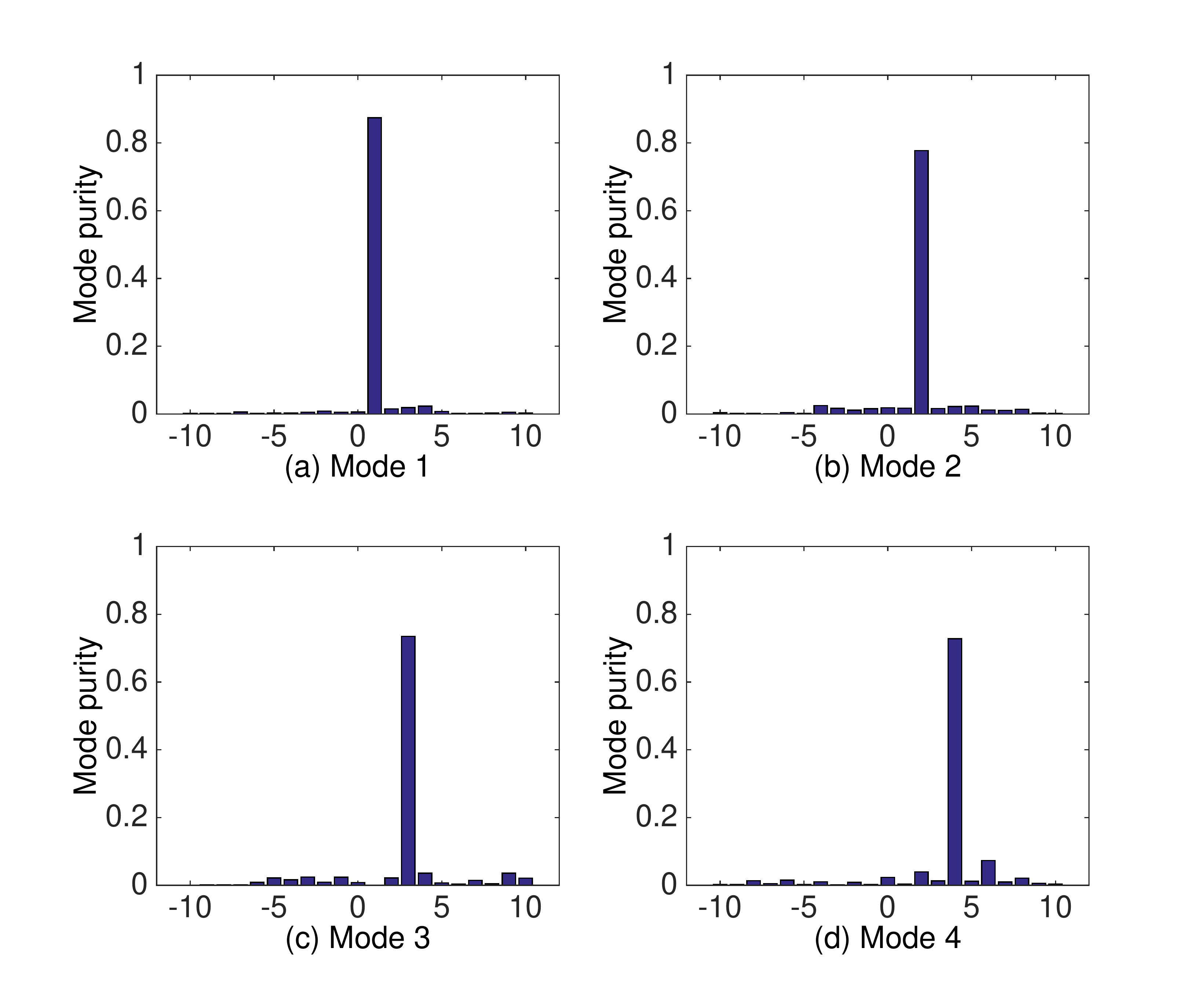}
 \caption{Mode purity of single mode at the position of $\theta=90^{o}$ (Simu): (a) $l=1$ (b) $l=2$ (c) $l=3$ (d) $l=4$}
 \label{mp}
\end{figure}

Fig.\ref{mp} shows the purity spectrum of each mode. It can be seen that all the generated PSOAM beams have high mode purity.
 
 \begin{table}[htb]
\setlength{\belowcaptionskip}{0.3cm}
\centering
\caption{THE MLD AND RP OF SINGLE MODE AND STRUCTURAL PATTERN USING 8 MODES }
\label{t0}       
\begin{tabular}{l|lllll}
\hline\noalign{\smallskip}
$Mode$ &1 &2 &3 &4 & $8 modes$ \\
\noalign{\smallskip}\hline\noalign{\smallskip}
MLD(Simu)& $86.0^{o}$ & $93.0^{o}$& $95.0^{o}$ & $102.0^{o}$& $90.0^{o}$ \\
MLD(Meas)& $84.8^{o}$ & $87.5^{o}$& $91.7^{o}$ & $95.0^{o}$& $89.7^{o}$ \\
RP(Simu)& -0.24dB& -0.09dB& -0.22dB & -1.26dB&0dB \\
RP(Meas)& -0.38dB & -0.13dB& -0.22dB& -0.77dB& -0.05dB\\
\noalign{\smallskip}\hline
\end{tabular}
\end{table}
 
\begin{figure}[htb]
 \centering
 \includegraphics[width=3.5in]{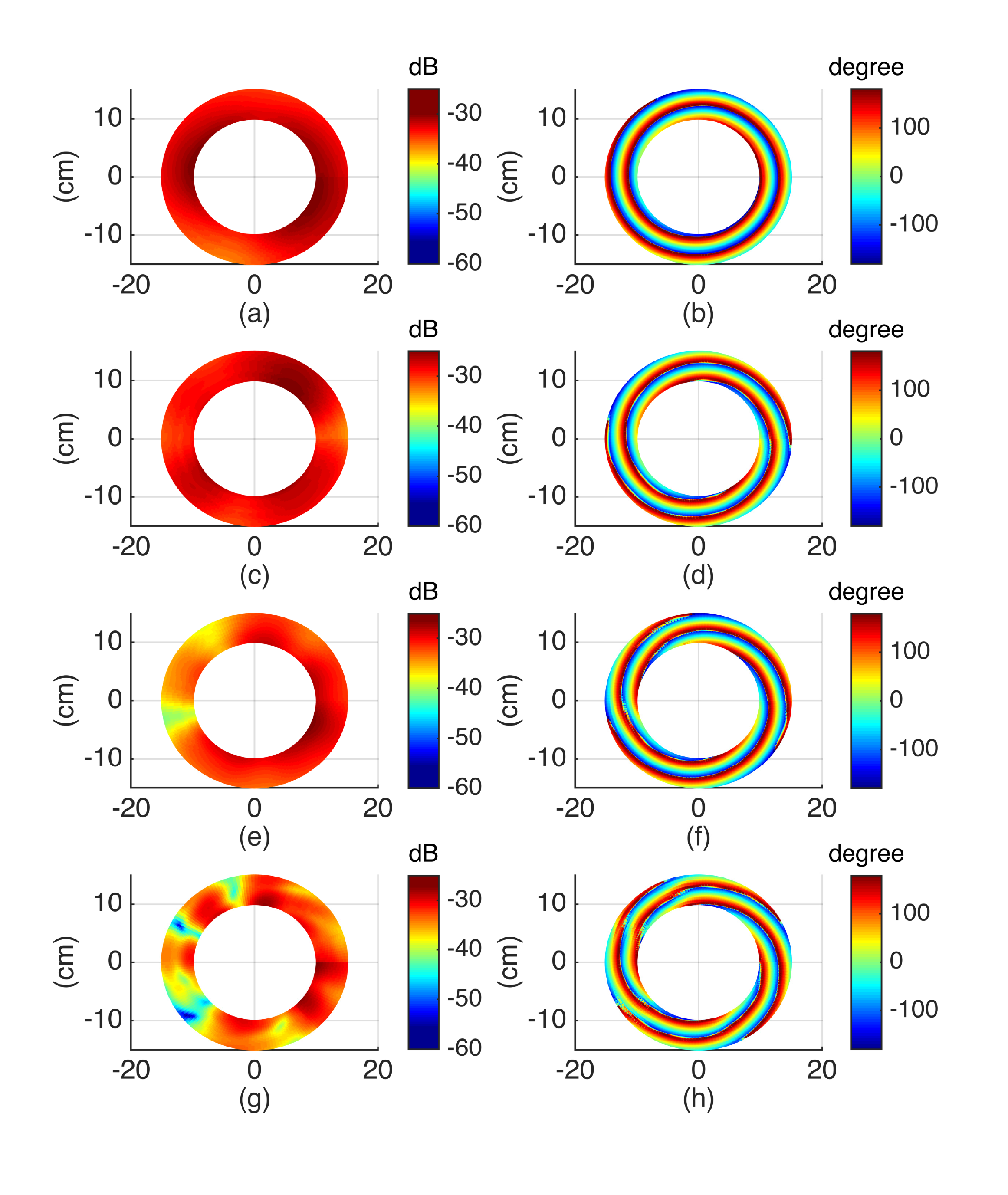}
 \caption{Power and phase distribution of single mode at $xoy$ plane (Meas): (a) power of $l=1$ (b) phase of $l=1$ (c) power of $l=2$ (d) phase of $l=2$ (e) power of $l=3$ (f) phase of $l=3$ (g) power of $l=4$ (h) phase of $l=4$}
 \label{single2}
\end{figure}

Fig.\ref{single2} shows the measured results of the power and phase distribution of each mode at $xoy$ plane. The observation plane is a ring with a radius from $10$ $cm$ to $15$ $cm$. The phase distribution of each mode shows the vortex characteristic which correlates with the mode number. When the OAM mode increases, the homogeneity of the power decreases. The reason lies in the fact that , the resonator's diameter increases with the increasing of the OAM mode, and thus the feeding nonconformity aggravates.

\subsection{Pattern Reconfiguration}
For the designed antenna, the eight PSOAM waves can be generated independently. Hence, this antenna can be used to the pattern reconfiguration application \cite{Zheng2018Realization}. Any azimuthal target pattern can be reconfigured by calculating the complex OAM mode purity $\vec A_{l}$ and feeding the corresponding amplitude and phase to the mode $l$

\begin{equation}
\label{fx}
   \vec A_{l}=|A_{l}| e^{j\Delta \phi _{l}}=\int^{2\pi}_{0} \vec F(r,\theta)e^{jl\varphi}d\varphi
\end{equation}

\begin{figure}[htb]
 \centering
 \includegraphics[width=3.0in]{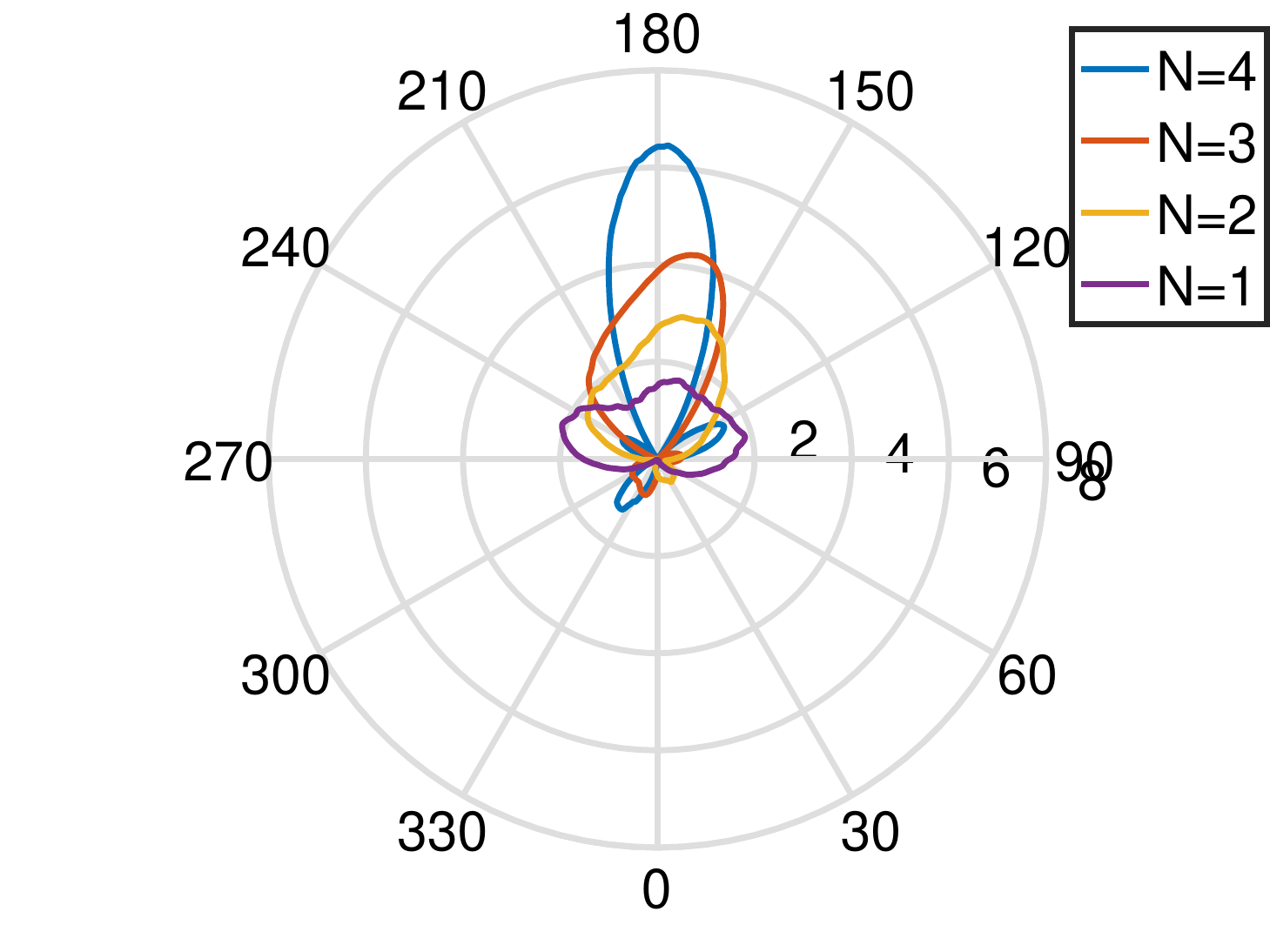}
 \caption{Azimuthal gain of configured beam with different mode numbers $N$ at the position of $\theta=90^{o}$ (Meas)}
 \label{1t4}
\end{figure}

Pencil beam can be constructed by multiply OAM beams with continuous mode of equal amplitude and same phase. Fig.\ref{1t4} shows three azimuthal pattern generated by the combinations of the PSOAM waves. It can be seen that with the increasing number of the superposed OAM waves $N$, the constructed beam becomes narrower. Theoretically the gain of the beam increases with the number of the OAM waves. The trend of the measured gain agrees well with the theoretical one.

\begin{figure}[htb]
 \centering
 \includegraphics[width=3.5in]{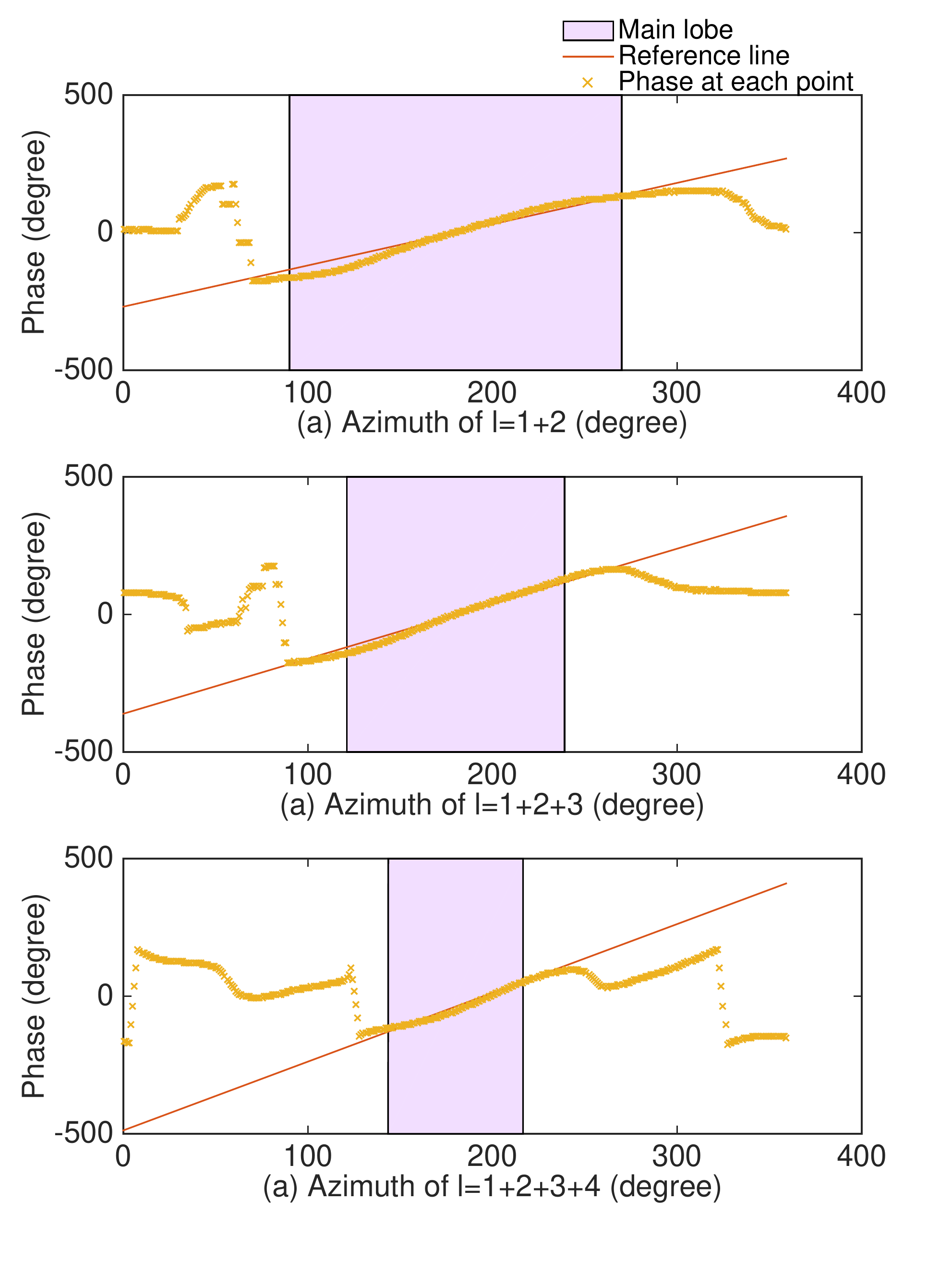}
 \caption{Azimuthal phase of configured beam with different mode numbers $N$ at the position of $\theta=90^{o}$ (Meas)}
 \label{1t4p}
\end{figure}

 Fig.\ref{1t4p} shows the phased distribution of the superposed beam within its main lobe at the plane of $\theta=90^{o}$. The slope of the three reference lines are the arithmetic mean of the superposed modes, which are $1.5,2,2.5$ respectively for the three cases. The results show that the phase slope in main lobe equals to the arithmetic mean approximately for each case, which means that this kind of pattern reconfiguration can maintain the vortex characteristics in its main lobe. 
 
\begin{figure}[htb]
 \centering
 \includegraphics[width=3.0in]{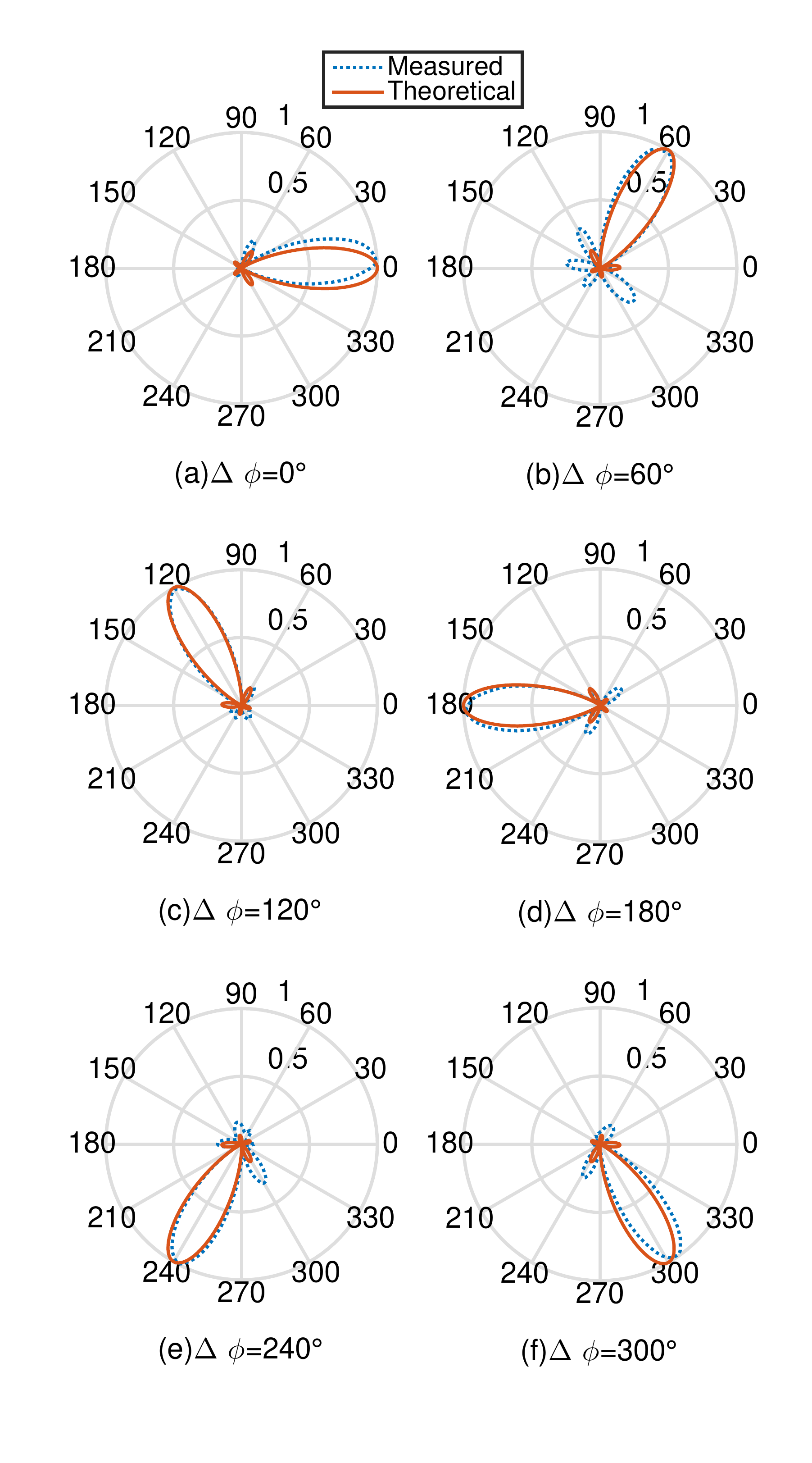}
 \caption{Normalized pattern of configured pencil beam configured by eight modes and its scanning (Meas)}
 \label{mode8sao}
\end{figure}

The measured results in Fig.\ref{1t4} and Fig.\ref{1t4p} prove that the pencil beam superposed by multi PSOAM waves mains the vortex characteristic of conventional OAM waves while increasing the directivity, which will have potential in the wireless communication. Fig.\ref{mode8sao} shows the pencil beam configured by all the eight modes whose initial phase are $0^{o}$, and its azimuthal scanning with MLDs at $\Delta \phi=0^{o}, 60^{o}, 120^{o}, 180^{o}, 240^{o}, 300^{o}$. The main lobe width (MLW) and MLD of different scanning angles are shown in Tab.\ref{t1}. The measured results show that the error of the actual MLD and the preset angle is small, which is mainly caused by the phase instability of the measurement system. The measured MLWs, whose theoretical value is $33.4^{o}$, changed little during the scanning, so the PSOAM based beam scanning has the ability of omni-directional $360^{o}$ scanning without distortion in azimuthal domain as expected.

\begin{figure}[htb]
 \centering
 \includegraphics[width=3.0in]{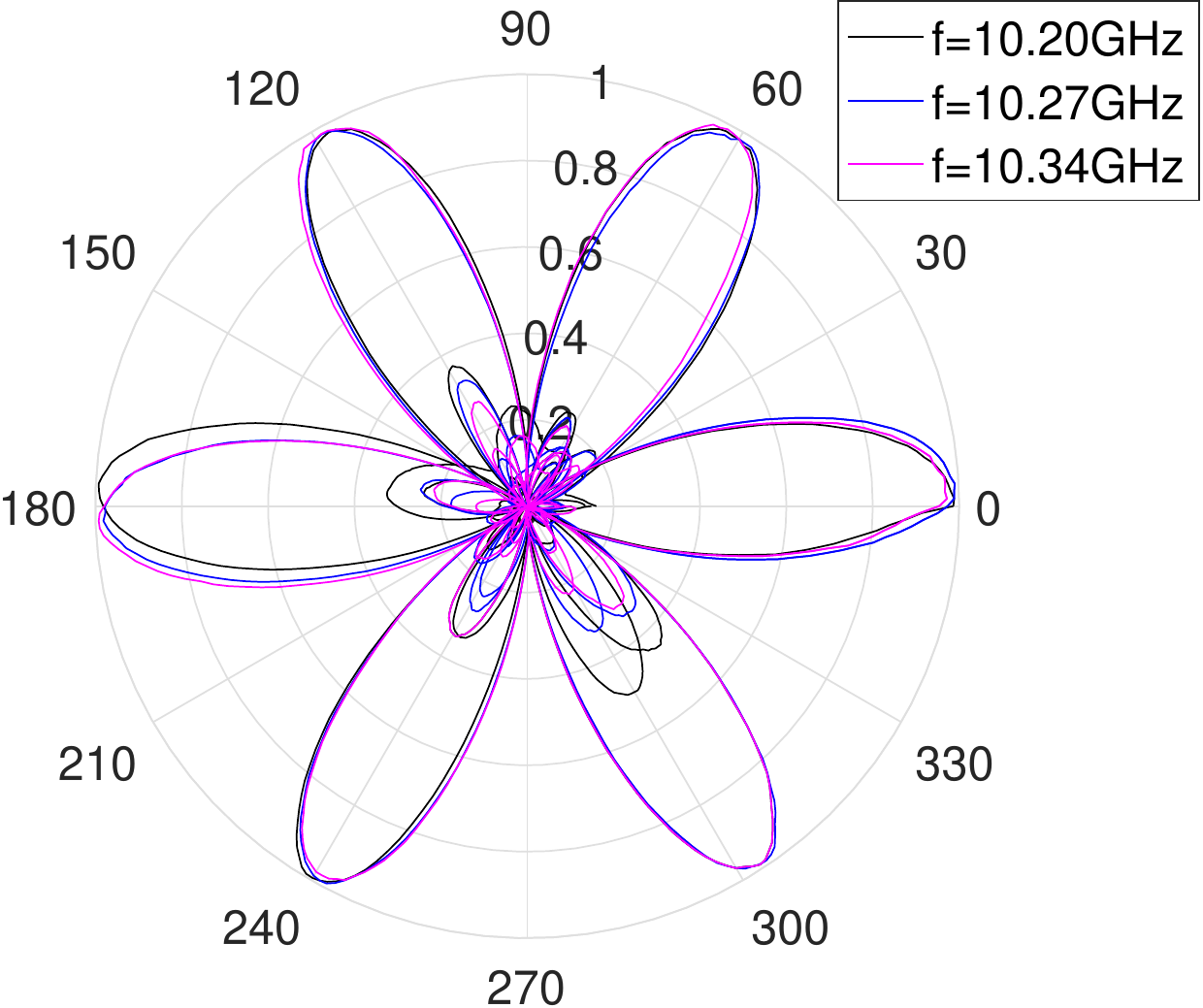}
 \caption{Scanning pattern of pencil beam at different frequency (Meas)}
 \label{pltx}
\end{figure}

In order to verify the stability of the configured pattern within its impedance bandwidth, Fig.\ref{pltx} shows the scanning pattern of pencil beam at the edge frequencies and center frequency of the bandwidth. The main lobes of the reconfigured patterns within the working bandwidth do not have obvious distortion. The change of frequency only affects the intensity of the side lobes, but not their main lobes, which means that beam scanning using the antenna is robust at the whole bandwidth. 

\begin{table}
\setlength{\belowcaptionskip}{0.3cm}
\centering
\caption{THE BEAM MLD AND MLW AT DIFFRERENT $\Delta \phi$}
\label{t1}       
\begin{tabular}{l|llllll}
\hline\noalign{\smallskip}
$\Delta \phi$ &$0^{o}$ & $60^{o}$ & $120^{o}$& $180^{o}$ & $240^{o}$& $300^{o}$ \\
\noalign{\smallskip}\hline\noalign{\smallskip}
MLD(Meas) & $1^{o}$ & $58^{o}$& $119^{o}$ & $182^{o}$& $242^{o}$ & $305^{o}$\\
MLW(Meas) & $36.0^{o}$ & $36.5^{o}$& $35.0^{o}$ & $36.0^{o}$& $35.0^{o}$ & $34.0^{o}$\\

\noalign{\smallskip}\hline
\end{tabular}
\end{table}
In conclusion, the PSOAM beams generated by this antenna can be used to configure pencil beam with any direction successfully. In fact, it can also be used to reconfigure other patterns of engineering significance in azimuthal domain, such as flat top pattern, multi-directional pattern and so on. In order to reconfigure pattern with less deviation, multimode PSOAM beam with more mode components and topological charge is necessary.

\section{Conclusion}

In this paper, a multimode PSOAM antenna, consisting of a CRG and a RPR, working in X-band with a bandwidth of $140 MHz$ is proposed. The CRG consists of four radially arranged top slotted TE coaxial resonators, which can generate eight modes conventional OAM-carrying beams independently. With the help of RPR, the MLD of OAM beams turn to the same plane of $\theta=90^{o}$, solving the stack difficulty. This antenna can generate eight OAM beams with high mode purity. It is a novel compact antenna which can generate more than 8 PSOAM modes. Using this antenna, various pattern can be constructed and the azimuthal 360 beam scanning of 8 PSOAM superposed beam is experimentally demonstrated. The proposed antenna provides a viable solution to the generation of the multimode PSOAM beams, and makes the PSOAM based pattern reconfiguration possible, which will have more potential in OAM based applications such as wireless communication and radar sensing.  

%

%
%

\section*{Acknowledgment}

The authors acknowledge the funding supported from the National Natural Science Foundation of China under Grantnumber 61571391.

\ifCLASSOPTIONcaptionsoff
  \newpage
\fi
\bibliographystyle{IEEEtran}
\bibliography{ref}

\end{document}